\documentclass[balance,upint,subscriptcorrection,varvw,mathalfa=cal=boondoxo,spanish,french,vietnamese,russian,greek,pdf-a,colorlinks]{asmeconf}
\usepackage{adjustbox}
\usepackage{lipsum}
\usepackage{xcolor}
\usepackage{stfloats}
\usepackage{soul}
\newcommand\blfootnote[1]{%
  \begingroup
  \renewcommand\thefootnote{}\footnote{#1}%
  \addtocounter{footnote}{-1}%
  \endgroup
}
\hypersetup{%
	pdfauthor={Evan Taylor},				
	pdftitle={axioms and Guidelines for a Fidelity Quantification Approach},                  % <=== change to YOUR pdf file title
	pdfkeywords={modeling, fidelity, simulation, accuracy},% <=== change to YOUR pdf keywords
	pdfsubject = {Provides a information based approach to the concept of Fidelity and the metrics used to measure it},			  % <=== change to YOUR subject
}

\begin{document}

% Change these fields to the right content for your conference.
% You can comment these out if for some reason you don't want a header.
% Use title case (first letters capitalized), not all capitals

\ConfName{PREPRINT-Accepted for Proceedings of the ASME \linebreak 2025
International Design Engineering Technical Conferences and \linebreak 
Computers and Information in Engineering Conference}
\ConfAcronym{IDETC/CIE2025}
\ConfDate{August 17–20, 2025} % update 
\ConfCity{Anaheim, CA} % update 
\PaperNo{DETC2025-168449}
\title{PREPRINT: Axioms for Model Fidelity Evaluation} % <=== replace with YOUR title

\SetAuthors{%
	Evan Taylor\affil{1}
	Edward Louis\affil{1},
    Gregory Mocko\affil{1}\CorrespondingAuthor{gmocko@clemson.edu},
	}

\SetAffiliation{1}{Clemson University, Clemson, SC}
%	Note: Luis and Maria are not real people.  Henry and Catherine have been dead for >450 years.

%	To switch from inline author names to gridded names, use the [grid] option.

\maketitle

\begin{abstract}
Digital engineering has transformed the design and development process. However, the utility of digital engineering is fundamentally dependent on the assumption that a simulation provides information consistent with reality. This relationship is described as model fidelity. Despite the widespread use of the term, existing definitions of model fidelity often lack formal rigor in practical application, which leaves ambiguity in how this similarity should be evaluated. This paper presents seven fundamental axioms to aid the development of future fidelity evaluation frameworks. An example of a ground vehicle model is used under an existing fidelity evaluation framework to observe the applicability of these axioms. In addition, these axioms are used as a reference point for considering future opportunities in future
work related to model fidelity.

\end{abstract}
\blfootnote{DISTRIBUTION STATEMENT A. Approved for public release; distribution is unlimited. OPSEC9536}
\begin{nomenclature}
\EntryHeading{Acronyms}
\entry{$CA$}{Critical Angle}
\entry{$SoI$}{System of Interest}
\entry{I}{Information}
\entry{M} Model 
\entry{$X$}{Input Space}
\entry{$Y$}{Output Space}
\end{nomenclature}
\hspace{1cm}
\section{Introduction}
Digital engineering practice has revolutionized the design process. Rather than spending countless hours building physical prototypes and developing standardized tests, prototyping can be done virtually with a fraction of the resources. As a result, many organizations, such as the Department of Defense, have developed initiatives to promote digital engineering practice to speed acquisition time \cite{DoD}. However, all digital engineering efforts rely on a crucial assumption:  the simulation yields information similar to reality. Without this assumption, the utility of digital engineering tasks is gone. As a result, understanding this relationship between a model and reality is a fundamental question in modeling and simulation and is termed model fidelity. Model fidelity refers to the degree to which the model produces the same outcomes as a physical system \cite{Kemple, Moon2013, glossary}. However, this conceptual definition of model fidelity leaves room for interpretation of how this similarity is measured. In this paper, seven axioms related to fidelity evaluation frameworks are presented. Although these axioms are not exhaustive, and prior authors have proposed axioms related to model fidelity, establishing a foundation for generalized fidelity evaluation frameworks is essential for advancing digital engineering as a discipline \cite{Roza2004, Schricker}.

\subsection{Fidelity Evaluation Frameworks}
Generally, a fidelity quantification system needs to evaluate both the qualitative and quantitative traits of a model, \( \mathcal{M} \). The qualitative traits of the model, which are codified through a set of information, \( I \), provide a mapping from the input space, \( X \), to the output space, \( Y \). A model can be defined as a mapping from the input space to the output space given some set of information I

\begin{equation}\label{eqn:1}
\mathcal{M} : X \times I \to Y.
\end{equation}

This means that for any given input \( x \in X \), the model produces an output \( y \) according to:

\begin{equation}\label{eqn:3}
y = \mathcal{M}(x, I).
\end{equation}

This model \( \mathcal{M} \) derives its value through its similarity to the system of interest. This similarity to the system of interest is broadly defined as model fidelity. However, this conceptual definition, as described in Section \ref{conceptual}, cannot be evaluated practically. As a result, a framework for evaluating the difference between the simulated SoI and the actual SoI must be defined. Generically, this can be treated as a function of the information contained in the model, \( I \), and the model's output, \( Y \):

\begin{equation}\label{eqn:2}
F : I \times Y \to [0,1].
\end{equation}

The fidelity evaluation function maps the qualitative and quantitative aspects of a model to a continuous range normalized between zero and one. This normalization is common practice within fidelity quantification frameworks as detailed in Section  \ref{binary} \cite{STATCOE,ClarkDuncan}.

Since the model produces outputs according to Equation \ref{eqn:3}, we can rewrite Equation \ref{eqn:2} as:

\begin{equation}\label{eqn:4}
F : I \times \mathcal{M}(X, I) \to [0,1].
\end{equation}

This generic formulation of a fidelity evaluation function, F, is aimed to encompass all possible functions. Quantitative and qualitative fidelity measures can be categorized as particular cases of this equation in which only one term is considered. Qualitative fidelity measures consist of a synthesis of the information contained within the model to judge the quality of the model. For example, Schricker et al. provide a fidelity evaluation framework of an elevator system model based on qualitative traits \cite{Schricker}. In contrast, quantitative fidelity measures rely upon the execution of the model. In literature, these types of fidelity evaluation frameworks are much more common. In these types of frameworks, statistics like accuracy and variability can be compressed into a fidelity score \cite{STATCOE}.

\subsection{Fidelity as a Information Based Metric}
\begin{figure}[h!]
    \centering
    \frame{\includegraphics[width=.85\linewidth]{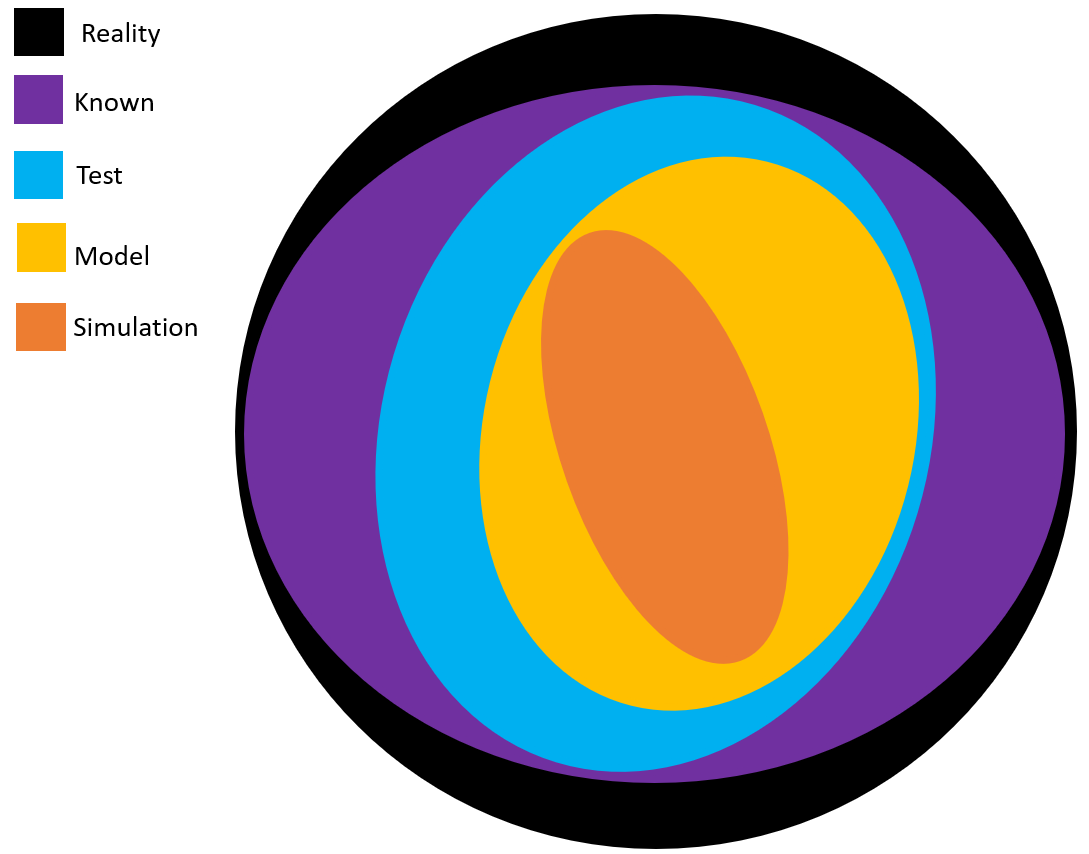}}
    \caption{Set Based Characterization of Model Fidelity}
    \label{fig:cat}
\end{figure}

Previous work by the authors presented a set-based framework to describe model fidelity, driven by the necessity for a stronger, information-based definition. The model development process systematically discards real-world information throughout the process. This process of model development can be decomposed and described as several states of knowledge: known phenomena, testing, modeling, and simulation—each of which contributed to a cumulative loss of fidelity. A graphical representation of this set-based approach is found in Figure \ref{fig:cat}.

This approach defined fidelity as a function of retained information from reality in the model at the time of simulation \cite{Taylor2024}. It is this retained information about the system of interest that provides value in a model. Models and simulations preserve observed information so that it can be used later to enable decision-making. Other authors have adopted similar approaches to model evaluation and validation, such as Hazelrigg's value-of-information approach to model validation. However, this does not encompass all perspectives on how fidelity should be conceptualized and evaluated. As a result, the following axioms are formalized to provide a general framework for fidelity quantification.

\section{Axioms for Fidelity Quantification}
\subsection{ Subjective and Conceptual}\label{conceptual}
\begin{figure}[h!]
    \centering
    \includegraphics[width=\linewidth]{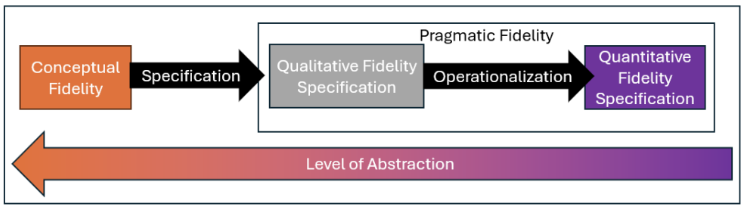}
    \caption{Graphical Description of the Relationship between Esoteric and Pragmatic Fidelity}
    \label{fig:qual_quant}
\end{figure}

\textbf{Axiom 1: Fidelity evaluation is necessarily subjective and conceptual in nature}\\
One of the fundamental issues of fidelity quantification is that fidelity is, by nature, a conceptual definition. In a dissertation on the topic, Roza's first axiom about fidelity quantification states that the "most correct formulation of simulation fidelity, 'esoteric fidelity' can never fully be articulated in practice \cite{Roza2004}.  In Roza's framework, esoteric fidelity is defined as the "inverse difference between reality and simulated reality". This provides the highest abstraction definition of fidelity from which others can be derived. However, this definition does not lend itself to practical use. In comparison, pragmatic fidelity definitions, which derive from this conceptual definition, provide the user methods for comparison and documentation. Figure \ref{fig:qual_quant} describes the relationship between these contrasting definitions of fidelity. 

This definition of esoteric fidelity requires the modeler to perform three tasks which require several subjective steps in order to achieve a formalized system: 

\begin{itemize}
\item Observe the system of interest 
\item Observe the simulated system
\item Determine the differences between the two. 
\end{itemize}

A commonly circulated phrase among the modeling and simulation community is the phrase "All models are wrong, but some are useful" \cite{Box_1976}. This wrongness is partially attributable to the subjectivity inherent in human observation. Models are based on a finite number of observations under a finite number of environmental conditions. In addition, models suffer from the biases of the observer, who decides which variables are controlled and measured \cite{Wimsatt_2006}. At best, modelers can strive for a sufficiently and conditionally correct model, but not a true representation. The conditional correctness of models is detailed more extensively in Section \ref{scenario}. 

Similar to the real-world system, a simulated system can only be evaluated a finite number of times, meaning that the modeler will choose a sampling method. In addition, continuous real-world phenomena must be discretized in order to perform simulation. In some domains, this may be one of the largest sources of uncertainty. For example, Gel et al. found that spatial discretization was the largest source of uncertainty in their computational fluid dynamics model 
 \cite{discrete}.

The method of comparison plays a critical role in fidelity evaluation and is a major source of subjectivity. One key decision in this process is how to represent both reality and the simulated system. While fidelity is defined in terms of the inverse difference between reality and simulation, this relationship is not strictly mathematical. However, it must maintain dimensional homogeneity, meaning that both representations describe the exact physical quantities. A comparison between items only makes sense if it is apples to apples, to use a common phrase. 

A related concept, commensurability, has been used to examine relationships between scientific theories \cite{sep-fleck}. Two theories are considered commensurable if they share a common framework that allows for meaningful comparison. Extending this idea to fidelity evaluation, both the system of interest and the models that claim to represent it must be commensurable with each other. This requirement has led to a heavy reliance on numerical accuracy as a primary measure of fidelity. Numerical measures of similarity provide a convenient methodology of comparison.

However, fidelity should not be solely determined by numerical accuracy. In physics-based modeling, the inclusion of relevant phenomena is often more critical than precise numerical agreement. This distinction gives rise to two broad approaches: quantitative fidelity evaluation, which emphasizes numerical accuracy, and qualitative fidelity evaluation, which focuses on the presence and representation of key phenomena. The choice and balance between these approaches is inherently subjective, requiring modelers to make judgment calls about how fidelity should be assessed and weighted in their evaluations.

Only when these subjective choices are made can the system actually be evaluated in terms of pragmatic fidelity. 

\subsection{Scenario Dependence}\label{scenario}
\textbf{Axiom 2: Fidelity can only be specified under a set of input conditions. }\\
Consider the following question: 
What is the fidelity of the small angle assumption? In other words, what is the fidelity of this model,  y=x, that claims to represent the system y=sin(x). This is graphically depicted below.

\begin{figure}[h!]
    \centering
    \frame{\includegraphics[width=\linewidth]{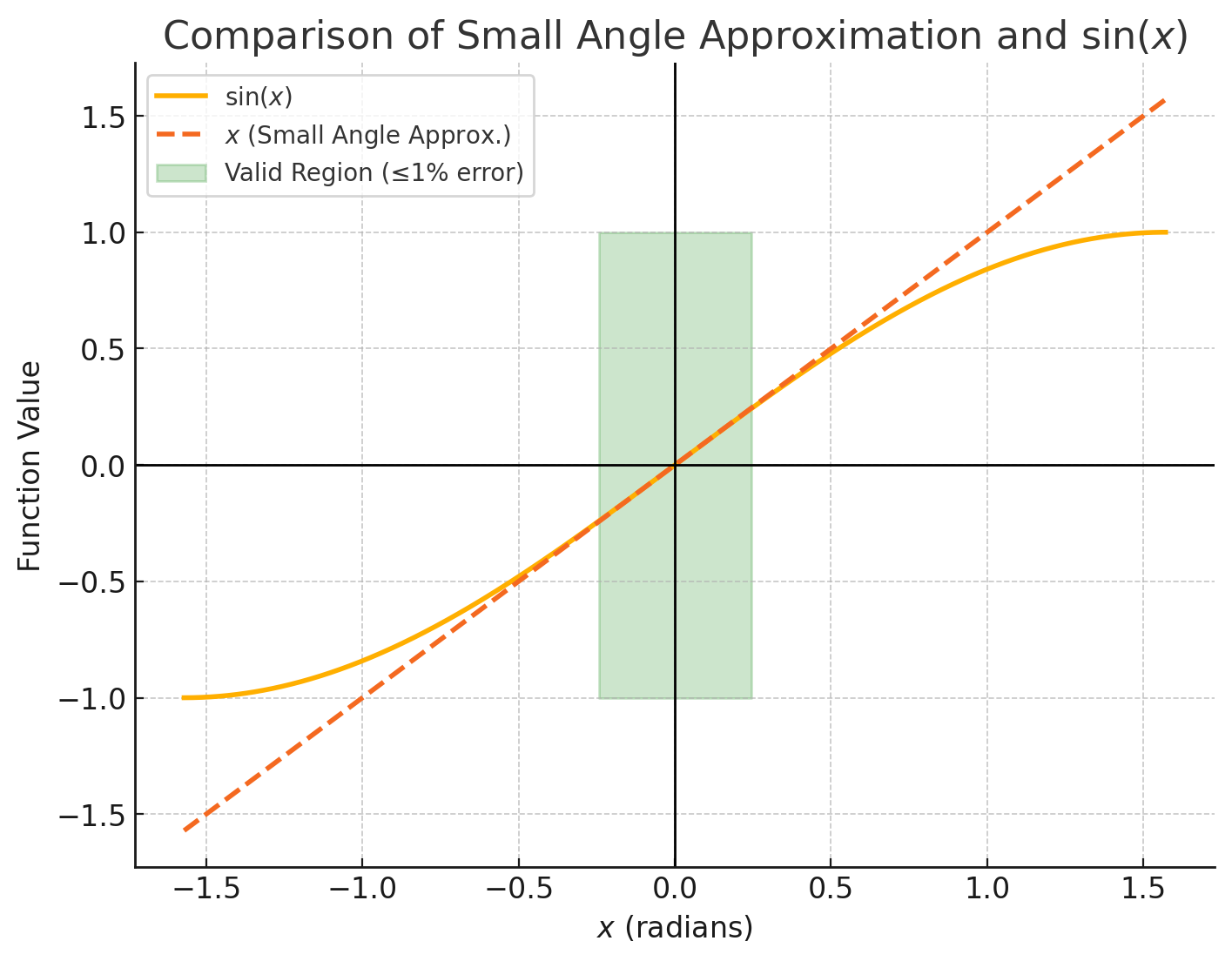}}
    \caption{Small Angle Assumption}
    \label{fig:enter-label}
\end{figure}

It is clear upon observation that the answer to this question requires specification of where the analysis should take place. An appropriate description of this model could be that, for small angles, the model matches reality, but this is not always the case. In this statement, it is clear that the model's relationship is directly related to the scenario. For the case above, a value of x, or a range of values of x, must be specified for the question. Once this is specified, along with a formalized method of comparison between the two, the model can be appropriately addressed. This could be done through qualitative or quantitative measures. For example, it could be said that the model represents reality if it is within a 1 percent error of the behavioral trace. In this case, subjectivity in the comparison framework is in the choice of one percent error. For a qualitative example, it could be said that the real system is locally linear in the area near x=0. In that case, the model mimics this behavior sufficiently for some range [$x_a$, $x_b$]. As a result, existing fidelity quantification frameworks either explicitly or implicitly feature an experimental frame under which the model shall be evaluated \cite{Ponnusamy, Roza2004}. 

\subsection{Referent Development}\label{referent}\hspace{1cm}
\textbf{Axiom 3: A model cannot be compared directly to the real world, instead the fidelity can only be evaluated relative to a best set of knowledge, called a referent, that represents the system of interest}\\
In Section \ref{conceptual}, it was described how models are intrinsically flawed, as human observation and testing are required to acquire data for modeling. However, even removing the subjectivity of human intervention, it may be impossible to capture all of the relevant phenomena for the SoI since there are an infinite number of phenomena that could influence the system. As a result, all analysis of a SoI must originate from a codified, finite set of information. This finite set of information can take the form of a test or a model that best represents the system of interest. 

\begin{figure}[h!]
    \centering
    \includegraphics[width=1\linewidth]{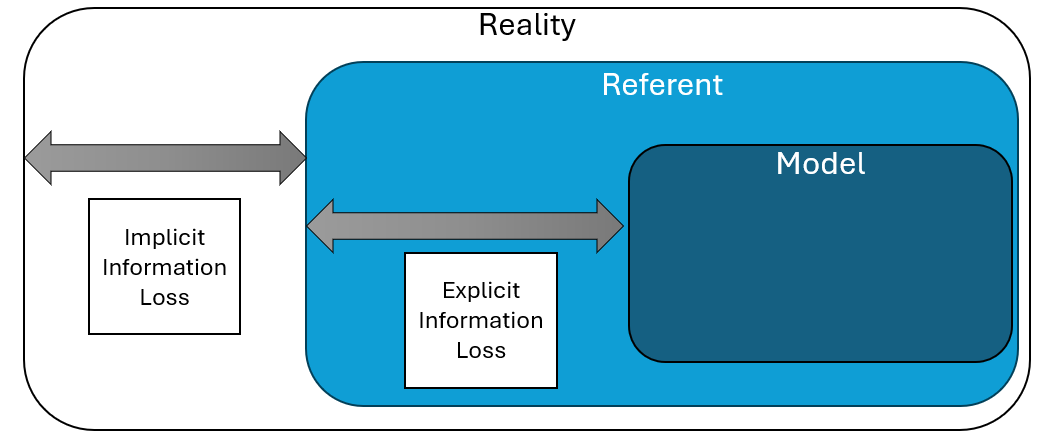}
    \caption{Graphical Representation of Information Content in Referent and Model}
    \label{fig:exp_imp}
\end{figure}

However, this knowledge conflicts with our semantic understanding of what model fidelity is supposed to represent, which is the level of similarity to reality. As a result, knowing the difference between our referent representation, under which we can judge fidelity, and the actual reality is a fundamental and necessary step in model management. Fundamentally, fidelity evaluation can only happen on the explicit, observed phenomena of the system rather than the unobserved phenomena in the real-world system. Figure \ref{fig:exp_imp} graphically demonstrates this relationship.

\subsection{Quantitative Methods}\label{Accuracy}
\textbf{Axiom 4: Accuracy may yield inaccurate conclusions about the information content of a model }\\
To elucidate how quantitative methods may mislead a modeler about the information contained within a model, consider the Gambler's paradox, as described as a coin toss experiment by Hazelrigg \cite{Hazelrigg2023}. A fair coin toss results in two equally likely outcomes: heads or tails. However, the result of an individual coin toss can be predicted using information about the toss itself, namely, the velocity and rotation. Suppose we have two models, A and B, which claim to represent the system of interest. Model A predicts the coin toss with a success rate of 50\% over a large number of observations, while Model B predicts the coin toss with a success rate of 20\%. Which of the two models, A or B, has the higher fidelity? Or, to rephrase, which of the two models has more information about reality? Counterintuitively, Model B, which is incorrect more often, would yield more information about the system of interest. A decision maker, with knowledge of Model B's relationship to reality, could use the inverse prediction to achieve a higher level of success. A similar issue arises when accuracy is the sole criterion for model evaluation, as seen in cases of overfitting.

\begin{figure}[h!] \centering \includegraphics[width=1\linewidth]{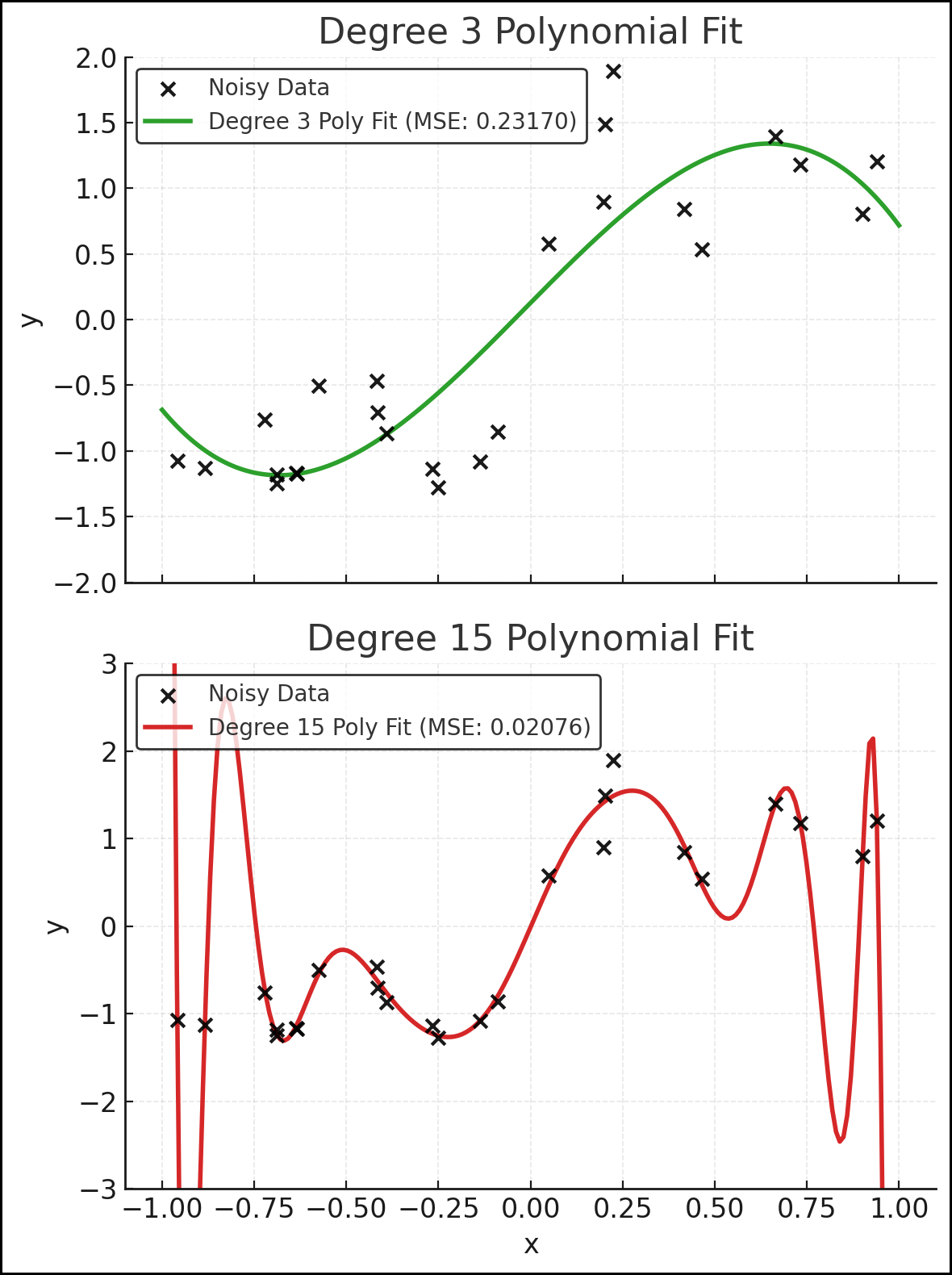} \caption{Comparison of two Polynomial Models} \label{fig:overfit} \end{figure} In Figure \ref{fig:overfit}, two candidate models are presented. Model A, a cubic polynomial, and Model B, a high-order polynomial, are fitted to a set of noisy data. From an accuracy-based perspective, the choice is clear: Model B should be chosen because the error rate is much lower. However, this decision ignores the much more salient question: Which model is more likely to represent the system of interest? From that perspective, the choice of model is much less clear.

This is not to say that fidelity cannot be evaluated based on outputs, but rather that it requires careful consideration. Haim and Hemez claim that a model's credibility should be assessed through three factors: accuracy to observed data, robustness to unobserved data, and consistency with similar models \cite{Haim_Hemez}. As seen in the example above, over-reliance on accuracy to observed data has an antagonistic effect on robustness.\newpage

\subsection{Normalization of Score} \label{binary}

\textbf{Axiom 5: A fidelity quantification system must normalize scores to aid the interpretation of the results. }\\
Among the many definitions and frameworks in fidelity evaluation, one of the most widely accepted principles is that fidelity scores should be normalized, typically within a fixed range [0,1], to aid interpretability. This was formalized in one of the first fidelity quantification efforts by Freeman and Gross in 1997 \cite{gross1997measuring}. In their system, a fidelity value of 1 represents complete adherence to the referent (i.e., a fully faithful representation). In contrast, a fidelity value of 0 indicates a complete lack of correspondence with the referent. As discussed in Section \ref{Accuracy}, this may not mean a completely inaccurate response. A fidelity value of 0 does not necessarily indicate a completely inaccurate response but rather a response that lacks any systematic relationship to the referent, providing no usable information about the system. 

Another reason for normalizing fidelity scores is the common perception of model fidelity as a binary concept. The word "fidelity" originates from the Latin fidēlis, meaning faithful, loyal, or trustworthy \cite{Merriam-Webster}. These qualities are often framed in opposition to their antonyms—unfaithful, disloyal, and untrustworthy. This binary interpretation has historically influenced fidelity assessment in digital engineering, where models are often categorized as either "sufficiently representative" or "insufficient" rather than evaluated along a continuous scale. Early fidelity quantification efforts reflected this perspective. Clark and Duncan propose a subjective rating system in which the modeler rates how well the simulation matches varying conditions from 0 to 1, where 0 represents no consideration of the condition and 1 is a full representation \cite{ClarkDuncan}.

In multi-fidelity modeling, models are often categorized as high- or low-fidelity, depending on their computational cost and accuracy \cite{Godino_2023}. Medium-fidelity models, while theoretically bridging these extremes, are less commonly leveraged because they may lack the efficiency of low-fidelity models while not providing the full trustworthiness of high-fidelity models. Trustworthiness is only valuable if it is available in spades.  

\subsection{Surrogate Models}
\textbf{Axiom 6: The fidelity of a surrogate model is strictly less than the fidelity of the model used to generate it} \\
Surrogate models are low computational cost models used to approximate more expensive models in engineering simulation \cite{surrogate}. Techniques  such as Response Surface Modeling (RSM), Kriging, and Support Vector Machines (SVM) reduce the computational burden while aiming to preserve key response characteristics of the original system. However, it is important to note that a surrogate model only emulates the system of interest in a black-box manner \cite{surrogate2}. Since a surrogate model is trained using data generated from a referent model, its fidelity is inherently constrained by the accuracy and representativeness of that data. It cannot introduce additional information into the system beyond what is embedded in the sampled input-output pairs.

A surrogate model is a function that maps input parameters to a vector of quantities of interest (QoI). If well-developed, the surrogate aims to emulate the responses of a referent, \( R \), which generates the "true" outputs of the system of interest. A surrogate model \( \widetilde{\mathcal{M}} \) is an approximation of the referent model \( \mathcal{M} \) developed with a finite number of input-output pairs \cite{surrogate2}. Formally, this relationship can be expressed as:

\begin{equation}
x \in \widetilde{X} \mapsto \widetilde{y} = \widetilde{\mathcal{M}}(x), \quad \text{where } \widetilde{X} \subset X 
\end{equation}
such that:
\begin{equation}
\widetilde{\mathcal{M}}(x) \approx \mathcal{M}(x)
\end{equation}

The surrogate model is built from a finite set of data points, which introduces interpolation error for unsampled points. This error cannot be fully eliminated, as there are infinitely many possible sampling points in any continuous, real-numbered interval. The error of a surrogate model is tied to the density of sampling points. For example, a polynomial surrogate model's error is bounded by the following expression \cite{Cheney_Kincaid_2013}.

Let \( f \) be a function such that \( f^{(n+1)} \) is continuous on \([a, b]\) and satisfies \( |f^{(n+1)}(x)| \leq M \). Let \( p(x) \) be the polynomial of degree \( n \) that interpolates \( f(x) \) at \( n + 1 \) equally spaced nodes in \([a, b]\), including the endpoints. Then on \([a, b]\):
\begin{equation}
| f(x) - p(x) | \leq \frac{1}{4(n + 1)} M h^{n+1}
\end{equation}
where \( h = \frac{b - a}{n} \) is the spacing between nodes.

With denser sampling, the error can be minimized. However, even if interpolation error is minimized, the surrogate still lacks the deeper physical relationships embedded in the referent model, contributing to a fundamental limit on fidelity. Because surrogate models are constructed from a limited set of input-output data rather than from first principles or empirical observations, they inherently discard some level of physical or structural detail. 

In summary, both the quantitative and qualitative outputs of the surrogate model differ from those of the referent model in ways that yield less information about the system of interest.

\subsection{Invariance of Fidelity Comparison}
\textbf{Axiom 7: Relative and absolute fidelity comparisons should result in the same model selection}\\
There are two primary uses for fidelity quantification: analyzing the "trustworthiness" of a model before use and for a basis of comparison for model selection. These two methods of fidelity quantification can be referred to as absolute and relative fidelity of the model \cite{Taylor2024}. Absolute fidelity comparisons are relative to a referent body of knowledge, which constitutes the modeler's best knowledge of the system \cite{Schricker}. Relative fidelity comparisons are between models.

It can be shown that absolute fidelity comparisons should converge into the same result as relative fidelity comparisons. Consider models A and B, which represent reality in scenario R. The fidelity of those models is the inverse difference between reality and each model. However, this relationship does not directly translate mathematically. Consider a literal interpretation of this, shown in equation \ref{eqn:fid1}.
\begin{equation}\label{eqn:fid1}
F_a = (R-M_A)^{-1} 
\end{equation}
In this formulation of fidelity, the fidelity is captured as the inverse difference between referent, R, and Model A, $M_a$. If the model used is the referent itself, this results in an undefined value for fidelity, which provides computational challenges and violates axiom 4. Instead, fidelity can more accurately be characterized through a set-based approach, as done in previous work \cite{Taylor2024}. In this set-based framework, the inverse is treated as a set difference, resulting in equation \ref{eqn:fid2}.

\begin{equation}\label{eqn:fid2}
F_a= R-(M_a)= M_a
\end{equation}
In this framework, the fidelity of the model is the relative knowledge content of the model at the time of simulation, S. It is then trivial to show that absolute and relative comparisons converge on the same. 

Relative fidelity:
\begin{equation} F_{A|B} = M_A - M_B \end{equation}

Absolute fidelity:
\begin{equation} F_A - F_B = M_A - M_B \end{equation}

Because these two equations are equal, relative and absolute measures of fidelity comparison should yield the same model rankings. If reference frame choice yielded different rankings, it would be a sign of inconsistency in the measurement of fidelity. This would have enormous implications for digital engineering, the prospective nature of which means that ground truth is not always evident. However, because both methods ultimately quantify the same underlying information, ensuring this agreement in a fidelity evaluation framework ensures a consistent and trustworthy measure for model selection.

\section{Application to an Existing Fidelity Evaluation Framework}
To demonstrate how these axioms apply to fidelity evaluation frameworks, an existing fidelity evaluation framework is applied and analyzed using data from a set of ground vehicle models. The experiment conducted is a digital recreation of a gradeability test, a standardized test used to assess a vehicle's ability to climb a longitudinal slope \cite{grade}.
\begin{figure}[h]
    \centering
    \makebox[\linewidth]{ % Forces the images to stay in one horizontal line
        \begin{subfigure}{0.48\linewidth}
            \centering
            \includegraphics[width=4cm, height=3cm]{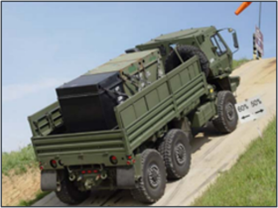}
            \caption{Real World Test}
            \label{fig:test1}
        \end{subfigure}
        \hfill
        \begin{subfigure}{0.48\linewidth}
            \centering
            \includegraphics[width=4cm, height=3cm]{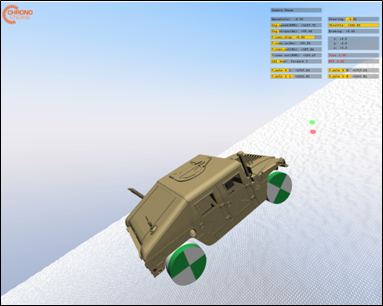}
            \caption{Chrono Test}
            \label{fig:test2}
        \end{subfigure}
    }
    \caption{Comparison between real-world and simulation }
    \label{fig:comparison}
\end{figure}

\begin{figure}[h!] \centering \includegraphics[width=1\linewidth]{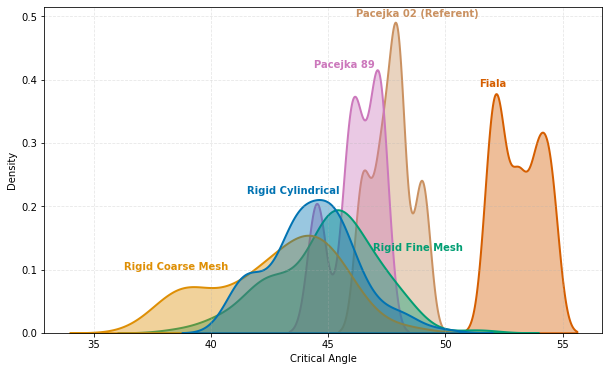} \caption{Distribution of Critical Angles Across Tire Models} \label{fig:results} \end{figure}

However, the fidelity evaluation framework under consideration is stochastic. As a result, uncertainty is introduced into the model in two forms. Aleatoric uncertainty is introduced using normally distributed soil parameters in the Bekker-Wang empirical model. An additional and unquantified source of uncertainty comes from random roughness of the driving surface. In each set of tests, the critical angle (CA)—the maximum slope the vehicle can climb—is determined using the bisection method for slopes ranging from 30\% to 70\% grade, with a resolution of 0.25\% grade \cite{Heister_Rebholz_2015}. This test is implemented in the open-source physics engine Chrono\cite{CHRONO}. The fidelity evaluation framework will be applied to vehicle models tested under these conditions, each incorporating different tire models. The results of this Monte Carlo simulation are shown in Figure \ref{fig:results}. 

The fidelity evaluation framework under study, developed by Weeks et al. \cite{STATCOE}, is a quantitative framework that infers the information content of a model. In this framework, a model is considered high fidelity if its output distribution closely resembles the referent distribution. 
This fidelity evaluation framework is formalized by the following equation:
\begin{equation}
f = f_a f_v 
= e^{-\frac{1}{2}\left( \frac{\bar{x}_m - \bar{x}_r}{s^*_r} \right)^2} 
\, e^{-\frac{\left(s^*_m - s^*_r\right)^2}{s^*_m s^*_r}}
\end{equation}

The differences between the means of the model and the referent, $\bar{x}_m$ and $\bar{x}_r$, and the differences between the standard deviations, $s^*_m$ and $s^*_r$ are used to calculate a scalar fidelity score, $f$. While using these quantities to assess fidelity is a rational approach, it remains a subjective decision (Axiom 1). Consequently, violations of the underlying assumptions in this framework may lead to misleading quantitative assessments (Axiom 4).

The fidelity of a model in this framework is specified within an input range, though not explicitly. Instead, the input range is defined by the distributions of inputs used to generate the model output. In this case, it corresponds to the normal parameter distributions and terrain height variations under which the model's fidelity is assessed (Axiom 2). Additionally, the term "referent" is explicitly used in this framework to denote the best available knowledge against which the model is judged (Axiom 3). In this case, a subjective understanding of features involved in the Pajecka 02 tire model is used to establish this model as the referent. 

Axiom 5 is embedded in the formulation of the fidelity evaluation function. If a model's mean and standard deviation match those of the referent, the fidelity score is 1. Moreover, since the exponents in the evaluation function are always negative, the fidelity score is bounded within the interval (0,1].

A key assumption in this framework, that the model’s output follows a normal distribution, is often violated and is violated in this case. In this experiment, various tire models were evaluated under different ground conditions, with the Bekker-Wang empirical model incorporating normally distributed parameters and a randomly generated bumpy terrain. Additionally, a random forest algorithm, implemented using the sklearn Python package, was used to generate a surrogate model based on the outputs of the Pacejka 02 referent model.

\begin{table*}[t!] % 'b' puts it at the bottom, ! allows better float placement
    \centering
    \begin{tabular}{|l|c|c|c|c|c|c|}
        \hline
        \textbf{Tire Model} & $f$ & $f_a$ & $f_v$ & Percent Error & Mean CA & Standard Deviation CA \\
        \hline
        Rigid Cylindrical & 1.82E-04 & 3.50E-04 & 0.519 & 7.085 & 44.334 & 1.866 \\
        Rigid Coarse Mesh & 9.30E-09 & 3.77E-08 & 0.246 & 10.385 & 42.760 & 2.607 \\
        Rigid Fine Mesh & 1.56E-03 & 0.004 & 0.373 & 5.878 & 44.910 & 2.205 \\
        Fiala & 1.21E-09 & 1.22E-09 & 0.987 & 11.379 & 53.145 & 0.949 \\
        Pacejka 89 & 0.194 & 0.199 & 0.978 & 3.193 & 46.191 & 0.982 \\
        Random Forest Model & 0.380 & 1.000 & 0.380 & 0.014 & 47.708 & 0.328 \\
        Pacejka 02 (Referent) & 1.000 & 1.000 & 1.000 & 0.000 & 47.715 & 0.847 \\
        \hline
    \end{tabular}
    \caption{Comparison of Tire Models}
    \label{tab:tire_models}
\end{table*}

The fidelity values in Table \ref{tab:tire_models} are generally low because the mean values of the different models diverge significantly relative to their standard deviations. This highlights a common issue with fidelity quantification: calibration sensitivity. While the authors of the original framework presented a well-formulated fidelity quantification system, their study examined models with distributions much closer to one another. In contrast, the fidelity quantification system applied here is more discriminative against low-accuracy models, making differences between models more apparent.

Additionally, the surrogate model generated using the random forest algorithm exhibits a fidelity score lower than the referent used to generate it (Axiom 6). Regarding Axiom 7, this framework employs an absolute reference frame, meaning that fidelity is evaluated relative to the referent. However, if the fidelity assessment were instead based on comparing individual models’ means and standard deviations directly, the resulting model rankings would likely remain consistent.

An important consideration for any fidelity evaluation framework is the nature of the metric itself. Any time information is compressed, like reducing a model’s performance to a single fidelity score, some information is lost. As a result, the metric must be well-aligned with the needs of the modeler. In this case, the fidelity evaluation framework measures the similarity between the model’s output distribution and the referent’s distribution. If the goal is to get a reasonable estimate of vehicle performance under uncertain conditions, this may be an appropriate measure of the fidelity. However, if the objective is to select the best model for an individual prediction, this metric may not be sufficient. It may even be the case that a model has the opposite response to a change of parameters to the referent. As an example, the functions $y=x$ and $y=-x$ would produce identical distributions if subjected to a random parameter x, but vastly different individual results. 
\section{Opportunities for Future Fidelity Work}
The axioms proposed in this paper establish a foundation for future fidelity quantification research and, more broadly, for simulation studies employing an information-based definition of model fidelity. However, these axioms may also influence other areas of simulation research. Sections \ref{construct} and \ref{mfm} explore two specific ways in which these axioms present opportunities for further work in simulation.

\subsection{Constructed Models} \label{construct}
In modeling and simulation work, it is commonplace to construct a set of notional models to test a framework. An example of this is a novel optimization framework that considers computational resources in addition to a multi-fidelity optimization framework\cite{Tao2024-rp}. In order to provide a testbed for their proposed algorithm, they constructed a system of models to represent the overall system of interest. In their constructed model, for example, a high-fidelity model would represented by the equation $y=(x-1)^2$ and the low-fidelity model is represented by the equation $y=(x-1)^2 +0.3$. While this constructed model allowed the authors to validate their framework, this set of models does not follow the typical conception of fidelity. 

First, the low-fidelity model is actually more computationally expensive and mathematically complex than the high-fidelity model. However, this was dealt with in the paper by specifying a cost for each model.

Secondly, and more importantly, is the offset by 0.3. Multi-fidelity modeling only makes sense if both models claim to represent the system of interest. In this case, it is not clear that the low-fidelity model represents the same system of interest.

Finally, the accuracy of the model does not vary based on the input space. Instead, it is a static offset between the low-fidelity and high-fidelity models. As a result, the low-fidelity model and the high-fidelity model have similar qualitative traits over the entire input space. The maximum error of the low-fidelity model is a constant value for all use cases and scenarios. This may introduce issues in optimization, as both models will result in the same minima. In reality, the fidelity and accuracy of models depend on the input space under which they are given. In the case of multi-fidelity optimization, this introduces an interesting non-linearity of the system, where the current design points influence model choice and model choice influences design. This provides an opportunity and challenge for multi-fidelity frameworks such as the one described. 

\begin{figure}[ht]
    \centering
    \includegraphics[width=1\linewidth]{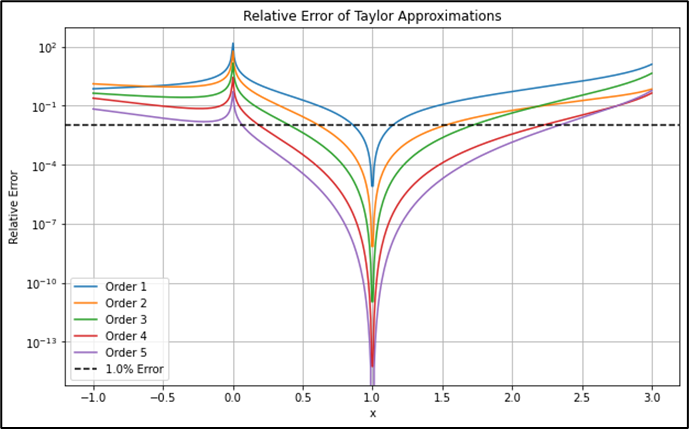}
    \caption{Set of Models Generated Using Taylor Series Approximations}
    \label{fig:Taylor}
\end{figure}

An example of how constructed models can be constructed using actual approximation methods, such as the Taylor series approximation at a point. 
\begin{multline}
f(x) = f(a) + f'(a)(x - a) + \frac{f''(a)}{2!} (x - a)^2\\ + \frac{f'''(a)}{3!} (x - a)^3 + \dots + \frac{f^{(n)}(a)}{n!} (x - a)^n + R_n(x)
\end{multline}
Consider the following set of approximations at x=1 for the function $y=sin(x)$ shown in Figure \ref{fig:Taylor}.

Each approximation provides a model of varying computational complexity and varying ranges of validity. All of the models are accurate about the operation point, but the range of validity grows as the approximation becomes more complex. In addition, all of these models are polynomials, which provide a continuous and differentiable approximation of the same approximated function. Astute readers may note that there is a sharp increase in error at x=0, which is where the behavior is locally linear. This locally linear behavior is captured in a previous example, the small angle approximation, which is a first-order Taylor approximation about x=0. 

\subsection{Multi-Fidelity Modeling} \label{mfm}
One of the most fruitful areas of research involving model fidelity is the area of multi-fidelity modeling. Computational time is an ever-present factor in digital engineering, and varying the level of fidelity can be a useful tool in speeding up this process. However, most multi-fidelity models only consider two models of the same phenomenon. However, many possible representations can be used as the low-fidelity model in this framework. In order to determine which model or model(s) should be used, the fact that fidelity is scenario-dependent can be leveraged. If the input space under which a variant model is high fidelity is known, then that model can be used in those circumstances. This concept can be shown using the following model of a helical coil spring that is part of a larger suspension model. .
\begin{figure}[ht]
    \centering
    \frame{\includegraphics[width=1\linewidth]{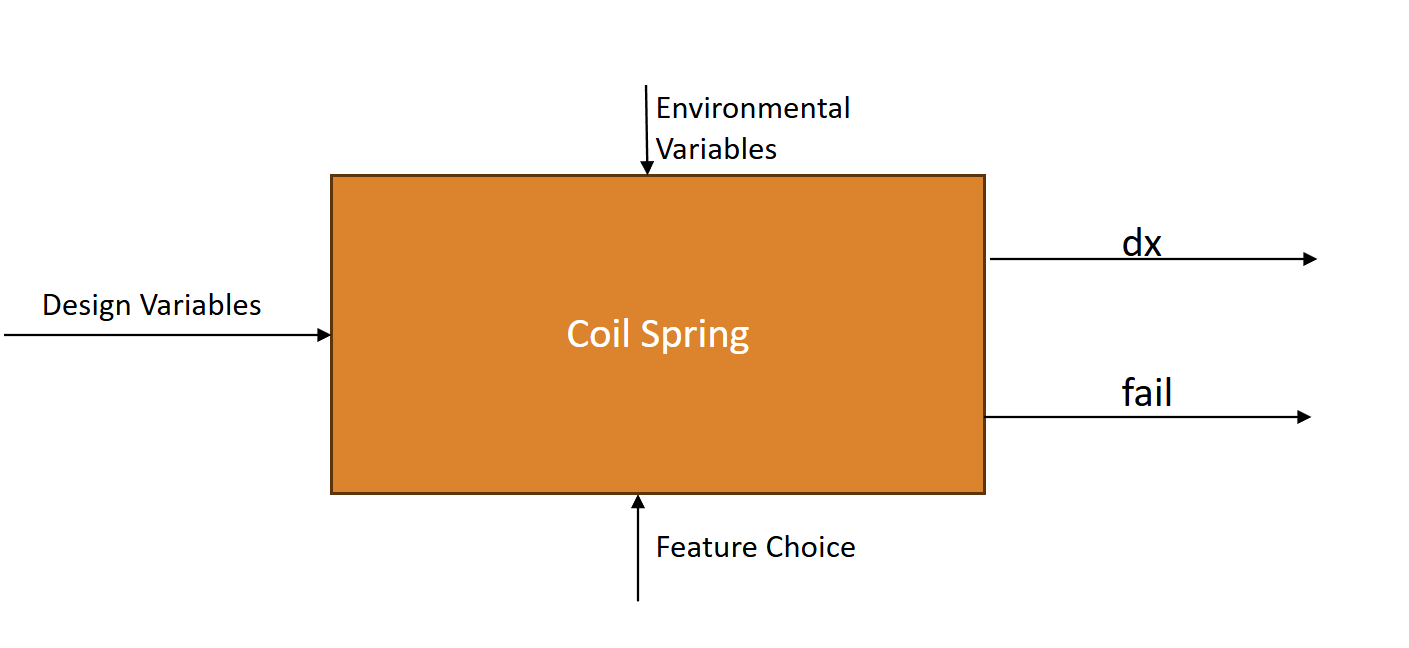}}
    \caption{Black Box Diagram of a Coil Spring Model}
    \label{fig:box}
\end{figure}

In this model, there are two primary outputs. The first is a total deformation of the spring, and the second is failure modes, in which the coil spring either breaks, bottoms out, or buckles. In addition, four features can be added or removed with a simple feature choice: temperature dependence, end condition of the spring, a buckling heuristic, and a buckling calculation. This is known as a 150\% modeling framework, in which variant models are encompassed within a single model \cite{Colletti2022-cx}. In this framework, a feature choice of [1, 1, 1, 1] would enable all of these features, which would result in the referent mod clearest example temperature dependence. A coil spring provides damping through compression, in which kinetic energy is dissipated through heat. As a result, there is a natural increase in the temperature of the coil spring through operation. As the temperature increases, the material properties change. As a result, the consideration of whether this feature should be included is related to the simulation scenario. However, this fact can also be used to select variant models. In the following design of experiments, each of the 16 possible model variants was run over a variety of input and design conditions. 

\begin{figure}[ht]
    \centering
    \begin{subfigure}{0.5\textwidth}
        \centering
        \includegraphics[width=\linewidth]{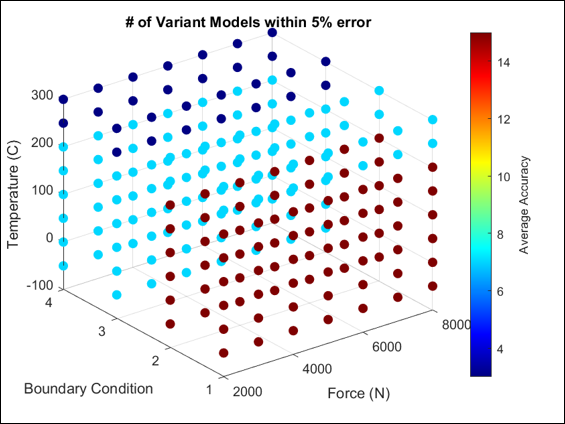}
        \caption{Coil Spring Variants within Deformation Accuracy Threshold}
        \label{fig:image1}
    \end{subfigure}
    \hfill
    \begin{subfigure}{0.5\textwidth}
        \centering
        \includegraphics[width=\linewidth]{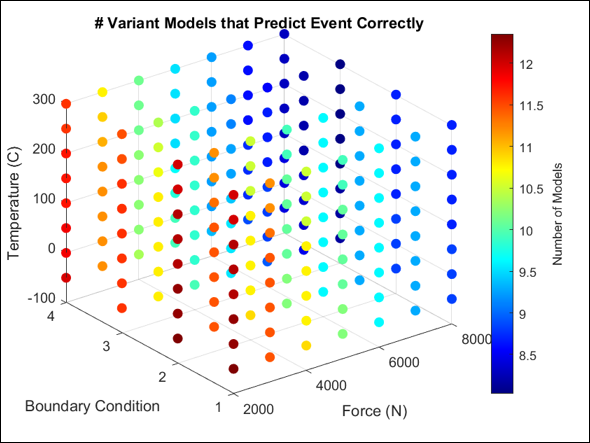}
        \caption{Coil Spring Variants that Predict Failure Correctly}
        \label{fig:image2}
    \end{subfigure}
    \caption{Demonstration of Feature Importance to Input Condition}
    \label{fig:side_by_side}
\end{figure}

After combining these two output conditions and finding the variant model that is the least computationally expensive, we can select the optimal model. If the location of these partitions was known a priori, this DoE was generated in 56.4\% of the time with acceptable results. In general, this formulation of multi-fidelity modeling represents a different view of how low-fidelity models should be used. Models are most valuable when they can neglect irrelevant information and can factor in relevant information. In a sense, this version of multi-fidelity modeling is a gray box view of the information contained within the model. Rather than viewing both models as black boxes with some relationship to one another, within this framework, the models contain varying levels of information according to the scenario of the simulation. While this methodology requires flexible programming methods and documentation, it allows for a more general approach to multi-fidelity modeling.

\newpage
\section{Conclusion}

This paper aims to establish a set of seven axioms for fidelity quantification using a arguments from literature and new contributions to the field. In order to emphasize the applicability
of these axioms, they are analyzed with respect to an existing fidelity evaluation framework and a set of ground vehicle models. In addition, these axioms are used to motivate future fidelity work, such as recommendations for construct models for analysis and multi-fidelity modeling. These axioms and the insights de-
rived from them aim to set standards for fidelity assessment and to enhance trust in simulation decision-making.

%%%%% Acknowledgments %%%%%%%%%%%%%%%%%%%%%%%%%%%

\section*{Acknowledgments}
This work was supported by the Virtual Prototyping Ground Systems (VIPR-GS) Center at
Clemson University and the Automotive Research Center (ARC), a US Army Center of Excellence
for modeling and simulation of ground vehicles, under Cooperative Agreement W56HZV-19-2-
0001 with the US Army DEVCOM Ground Vehicle Systems Center (GVSC). All opinions,
conclusions and findings wherein are those of the authors and may not be those of the
affiliated institutions. DISTRIBUTION STATEMENT A. Approved for public release; distribution is unlimited. OPSEC9536

\newpage
\bibliographystyle{asmeconf}  %% .bst file following ASME conference format. Do not change.
\bibliography{refs}%% <=== change this to name of your bib file

%%%  APPENDICES  %%%%%%%%%%%%%%%%%%%%%%%%%%%%%%%%
\appendix

\end{document}